\documentclass[pre,aps,twocollumns]{revtex4-2}
\usepackage{mathpazo}
\usepackage[T1]{fontenc}
\usepackage[latin9]{inputenc}
\usepackage{color}
\usepackage{amsmath}
\usepackage{amssymb}
\usepackage{graphicx}
\usepackage{subfigure}
\usepackage{ulem}
\usepackage[unicode=true,bookmarks=false,breaklinks=true,backref=false,colorlinks=true]{hyperref}
\hypersetup{citecolor=blue,filecolor=black,linkcolor=red,urlcolor=blue}

\begin{document}

\title{The Lugiato-Lefever equation driven by a double tightly focused  pump}
\author{Mateus C. P. dos Santos$^{1}$}
\author{Shatrughna Kumar$^{2}$}
\author{Wesley B. Cardoso$^{3}$}
\author{Boris A. Malomed$^{2,4}$}
\affiliation{$^{1}$Federal Institute of Maranhão, IFMA-PPGEM,
65030-005, São Luís, Maranhão, Brazil}
\affiliation{$^{2}$Department of Physical Electronics, School of Electrical Engineering,
Faculty of Engineering, and Center for Light-Matter Interaction, Tel
Aviv University, Tel Aviv 69978, Israel}
\affiliation{$^{3}$Instituto de Física, Universidade Federal de Goiás, 74.690-900, Goiânia,
Goiás, Brazil}
\affiliation{$^{4}$Instituto de Alta Investigación, Universidad de Tarapacá, Casilla
7D, Arica, Chile}

\begin{abstract}
We introduce a model of an optical cavity based on the one-dimensional
Lugiato-Lefever (LL) equation, which includes the pump represented by a
symmetric pair of tightly localized \textquotedblleft hot spots" (HSs) with
phase shift $\chi $ between them, and self-focusing or defocusing cubic
nonlinearity. Families of bound states, pinned to the double HS, are found
in the system's parameter space. They feature the effect of the symmetry
breaking (SB) between peaks pinned to individual HSs, provided that the
phase shift takes values $0<\chi <\pi $, and the LL equation includes the
loss term. The SB, which is explained analytically, takes place in the full
LL model and its linearized version alike. The same phenomenology is also
explored in the framework of the LL equation with the double HS and quintic
self-focusing. In that case, there are stable symmetric and asymmetric bound
states, in spite of the presence of the background instability driven by the
critical collapse.
\end{abstract}

\maketitle

\section{Introduction}

A fundamental property of nonlinear photonic systems, both conservative and
dissipative ones, is that the stable balance between the linear diffraction
and/or dispersion and nonlinear self-focusing may support many species of
self-trapped states in the form of solitons \cite{Kivshar_03,Peyrard}. In
dissipative systems, the existence of solitons also requires the balance
between the intrinsic loss and externally applied gain or pump \cite%
{Rosanov,Ferreira}. Detailed theoretical and experimental studies of
dissipative solitons have been carried out in various setups, in which the
background loss is balanced by gain, such as in lasing media. Such systems
are typically modelled by one- and two-dimensional (1D and 2D) equations of
the complex Ginzburg-Landau (CGL) type \cite{Grelu,2D-CGL}.

In externally driven nonlinear optical systems, the losses are compensated
by the pump in the form of illuminating laser beams. The fundamental model
of such passive systems is supplied by the Lugiato-Lefever (LL) equations
\cite{Lugiato_PRL87}. This class of models was theoretically analyzed in
many forms, including 1D and 2D ones \cite%
{Tlidi_C17,Panajotov_EPJB17,Tlidi_AOP22,Wabnitz}. They predict pattern
formation with diverse applications in nonlinear optics and laser physics
\cite{Coillet_PJ13}-\cite{Wabnitz}. Among these applications, extremely
important ones are the generation of Kerr solitons and frequency combs in
passive cavities \cite{Valcarcel_PRA13}-\cite{Dong_PRR21}. The generation of
terahertz radiation is another significant effect predicted by the LL
equation \cite{Huang_PRX17}.

Usually, LL models are considered with the spatially uniform pump, which is
appropriate for most experimental realizations. On the other hand, the
consideration of the LL equations with a localized pump is relevant too, as
it is an adequate model for passive optical cavities driven by tightly
focused laser beams. Note that strictly localized states can be predicted
only in the case of a localized pump, while the spatially uniform pump
necessarily gives rise to states existing on top of a nonzero background
field. In this context, analytical solutions of the 1D LL equations with the
focused pump represented by a delta-function of coordinate $x$
(\textquotedblleft hot spot", HS) were obtained in Ref. \cite{Cardoso_SR17}
(the tightly localized gain term $\sim \delta (x)$ makes it also possible to
produce exact solutions for pinned modes in the framework of CGL equations
\cite{HK1,HK2}). The same work \cite{Cardoso_SR17} reported approximate
solutions of the 2D LL equation with the tightly-focused Gaussian-shaped 2D
pump.

Solutions for localized states that represent robust pixels with zero
background were obtained in the framework of the 2D LL equation \cite%
{Cardoso_EPJD17}, that included the spatially uniform pump, cubic
self-interaction, and a tightly-confining harmonic-oscillator potential
(which provided the full localization, in spite of the action of the uniform
pump). Recently, classes of stable 2D solutions were reported in the
framework of the LL equation with a vorticity-carrying spatially localized
pump \cite{Shatrughna}.

The next natural step in the studies of localized pumped modes, which is the
subject of the present work, is to consider the LL equation with a set of
two mutually symmetric HSs, with a phase shift $\chi $ between them, which
represents an experimentally relevant setup for the passive cavity driven by
a symmetric pair of tightly focused beams. This setup essentially expands
the variety of localized states produced by the LL model, such as double
pixels. In particular, the solutions demonstrate symmetry breaking (SB) of
the established states supported by the underlying symmetric HS pair,
provided that the phase shift takes values $0<\chi <\pi $, and the LL
equation includes the loss term. The SB takes place in the framework of both
the full LL equation and its linearized version, the basic features of the
SB phenomenology being explained here analytically. We also perform the
analysis of symmetric and asymmetric states pinned to the double HS in the
framework of the 1D\ LL equation \ with quintic self-focusing. In the latter
case, stable bound states are found in spite of the presence of the
underlying instability driven by the critical collapse \cite{Berge,Fibich}.

The presentation in the paper is arranged as follows. The main model, based
on the cubic LL equation with the double HS, is introduced in Section 2,
which also includes analytical results, such as the explanation of the SB
phenomenology. Numerical results are reported, in a systematic form, in
Section 3. The quintic-LL model with the double HS is addressed, by means of
numerical methods, in Section 4. The paper is concluded by Section 5.

\section{The model}

\subsection{The Lugiato-Lefever (LL) equation with the cubic nonlinearity
and a pair of hot spots (HSs)}

We consider the LL equation for complex amplitude $u\left( x,t\right) $ of
the optical field in the passive 1D cavity with the cubic nonlinearity (the
quintic nonlinearity is considered below) and a pair of spatially separated
HSs, which represent tightly focused pump beams in the form of
delta-functions \cite{HK1,HK2}):%
\begin{equation}
\frac{\partial u}{\partial t}=-\alpha u+\frac{i}{2}\frac{\partial ^{2}u}{%
\partial x^{2}}+i\sigma \left( \left\vert u\right\vert ^{2}-\eta ^{2}\right)
u+\epsilon \left[ e^{-i\chi /2}\delta \left( x+\frac{l}{2}\right) +e^{+i\chi
/2}\delta \left( x-\frac{l}{2}\right) \right] .  \label{LLE}
\end{equation}%
Here $\alpha >0$ is the loss parameter, $\sigma =+1$ and $-1$ corresponds,
respectively, to the self-focusing and defocusing Kerr (cubic) nonlinearity,
real $\epsilon >0$ is the strength of each pump beam, $l$ and $\chi $ are
the separation and phase shift between them, and parameter $\eta ^{2}$,
which defines the cavity mismatch, may be positive or negative. In the
experiment, phase shift $\chi $ can be readily imposed by passing one pump
beam through a phase plate. The case of $\eta ^{2}>0$ is favorable for the
formation of soliton-like states, as the respective version of Eq. (\ref{LLE}%
) is similar to the nonlinear Schrödinger (NLS) equation with the cubic
self-focusing nonlinearity.

The set of two delta-functions and SB in this context were previously
studied in the framework of the conservative model, based on the NLS\
equation \cite{Thawatchai}. Here, a new feature is phase-shift $\chi $
between the two HSs, which has no natural counterpart in the NLS and CGL
equations. The presence of the latter parameter leads to a new effect, in
the form of the symmetry breaking (SB) in stable stationary states supported
by the symmetric HS pair. Indeed, the symmetry of the solution with respect
to the coordinate reflection, $x\rightarrow -x$, makes it necessary to
replace $\chi $ by $-\chi $, which requires the application of the complex
conjugation to the time-independent version Eq. (\ref{LLE}). However, the
presence of the loss term $-\alpha u$ in the equation breaks the invariance
of the equation with respect to this transformation.

Stationary solutions of Eq. (\ref{LLE}) are characterized by values of the
total power (alias norm),%
\begin{eqnarray}
P &=&\int_{-\infty }^{+\infty }\left\vert u\left( x\right) \right\vert
^{2}dx\equiv P_{+}+P_{-},  \notag \\
P_{+} &=&\int_{0}^{+\infty }\left\vert u\left( x\right) \right\vert
^{2}dx,~P_{-}=\int_{-\infty }^{0}\left\vert u\left( x\right) \right\vert
^{2}dx.  \label{P1D}
\end{eqnarray}%
The main issue to consider here is the SB (symmetry breaking) of stationary
localized patterns pinned to the symmetric HS pair, which takes place in the
case of $P_{+}\neq P_{-}$, and is characterized by the respective parameter,%
\begin{equation}
\mathrm{SB}\equiv \frac{P_{+}-P_{-}}{P_{+}+P_{-}}.  \label{SSB}
\end{equation}

The numerical solution of Eq. (\ref{LLE}) was performed by means of the
fourth-order Runge-Kutta algorithm, based on the Fourier spectral method
\cite{Yang_10}, replacing the ideal delta-functions by the regularized ones,%
\begin{equation}
\tilde{\delta}(x)=\frac{1}{\sqrt{\pi }\xi }\exp \left( -\frac{x^{2}}{\xi ^{2}%
}\right) ,  \label{delta}
\end{equation}%
with $\xi $ small enough. The computations were performed with $\xi =0.05$

\begin{figure}[tb]
\centering \includegraphics[width=0.9\textwidth]{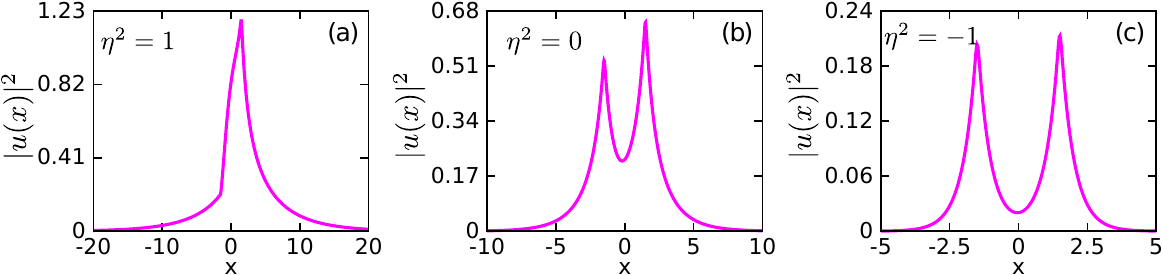}
\caption{Stationary profiles obtained numerically from Eq. (\ref{LLE}%
) for $\eta^2=1$, $0$, and $-1$. The other parameters are $%
\alpha=0.1$, $\epsilon=0.5$, $\chi=\pi/2$, $%
\sigma=-0.5$ and $l=3$.}
\label{Fig_et}
\end{figure}

\begin{figure}[tb]
\centering \includegraphics[width=0.9\textwidth]{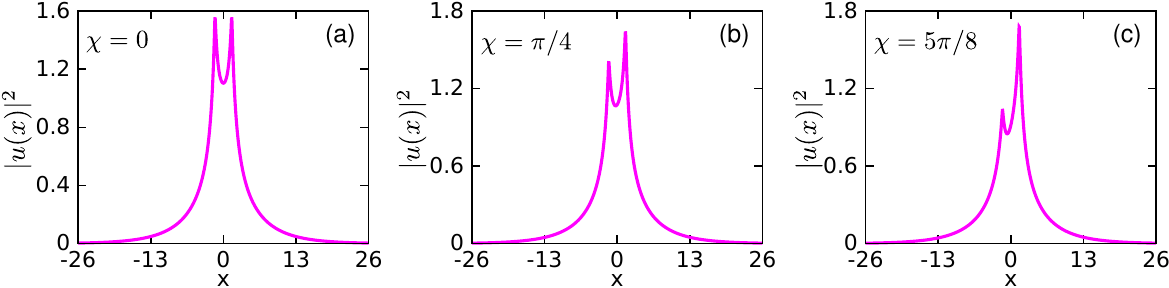}
\caption{Symmetric and asymmetric stationary profiles obtained numerically
from Eq. (\ref{LLE}) for different values of the phase shift $%
\chi $ between the two HSs. The other parameters are $\alpha %
=0.1$, $\epsilon =0.7$, $\eta ^{2}=1$, $\sigma =-0.5$
and $l=3$.}
\label{Fig_chi}
\end{figure}

\begin{figure}[tb]
\centering \includegraphics[width=0.45\columnwidth]{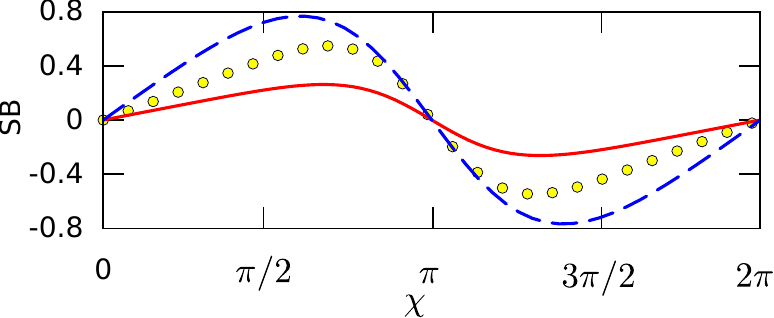}
\caption{The asymmetry ratio SB, defined as per Eq. (\ref{SSB}) vs. $%
\chi $, for the system with the self-repulsive sign of the
nonlinearity, $\sigma =-0.1$, $\epsilon =0.1$, $%
\alpha =0.1$, and $l=3$. The results represented by the dashed lines, yellow
circles, and red solid lines pertain to $\eta ^{2}=+1$, $0$ and $-1$%
, respectively.}
\label{Fig_SSB_chi}
\end{figure}

\begin{figure}[tb]
\centering \includegraphics[width=0.9\textwidth]{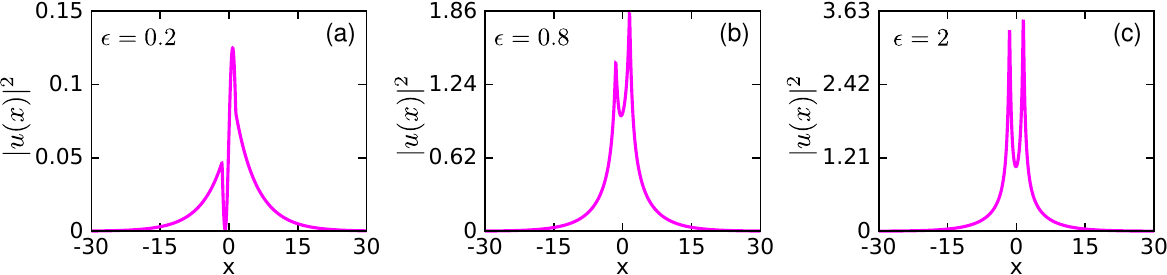}
\caption{Stationary profiles obtained numerically from Eq. (\ref{LLE}%
) for different values of the pump strength $\epsilon $. The other
parameters are $\alpha =0.1$, $\chi =\pi /2$, $%
\eta ^{2}=1$, $\sigma =-0.5$ and $l=3$.}
\label{Fig_ep}
\end{figure}

\subsection{Analytical approximations}

In the case of the self-focusing nonlinearity, $\sigma >0$, and $\eta ^{2}>0$%
, localized solutions pinned to the single delta-function pump can be
approximately found in the framework of the perturbation theory, assuming
that $\alpha $ and $\epsilon $ are small parameters. Setting, for this
purpose, $\sigma =\eta ^{2}=1$, the approximate solution can be taken as the
NLS soliton,
\begin{equation}
u_{\mathrm{sol}}\approx \sqrt{2}\exp \left( i\zeta \mp i\frac{\chi }{2}%
\right) \mathrm{sech}\left( \sqrt{2}\left( x\pm \frac{l}{2}\right) \right) ,
\label{sol}
\end{equation}%
with phase shift $\zeta $ determined by the power-balance condition between
the pump and loss:%
\begin{equation}
\frac{dP}{dt}=2\epsilon \cos \zeta \cdot A-2\alpha P=0,  \label{balance}
\end{equation}%
where $A\equiv \sqrt{2}$ is the amplitude of soliton (\ref{sol}), and $%
P\equiv 2\sqrt{2}$ is its power. Thus, Eq. (\ref{balance}) yields
\begin{equation}
\cos \zeta =2\alpha /\epsilon .  \label{cos}
\end{equation}%
Obviously, this solution exists under the condition of $\cos \zeta \leq 1$,
i.e.,
\begin{equation}
\epsilon \geq 2\alpha .  \label{thr}
\end{equation}

On the other hand, it is also possible to construct approximate
small-amplitude analytical solutions as those produced by the linearized
version of Eq. (\ref{LLE}). In the case of $\sigma \eta ^{2}>0$, the latter
equation with a small loss coefficient $\alpha $ admits the existence of
modes tightly pinned to the HS. Then, it is easy to find the respective
small-amplitude stationary solution of Eq. (\ref{LLE}) a vicinity of each HS
(provided that separation $l$ between them is very large):

\begin{equation}
u_{\mathrm{lin}}^{\mathrm{(tight)}}\approx -\frac{i\epsilon }{\sqrt{2\sigma }%
\eta }\exp \left( \mp i\frac{\chi }{2}-\sqrt{2\sigma \eta ^{2}}\left\vert
x\pm \frac{l}{2}\right\vert \right) ,  \label{lin}
\end{equation}%
with power%
\begin{equation}
P_{\mathrm{lin}}^{\mathrm{(tight)}}\approx \frac{\epsilon ^{2}}{\left(
2\sigma \eta ^{2}\right) ^{3/2}}  \label{Plin}
\end{equation}%
($\sigma $ and $\eta ^{2}$ are kept as free parameters here, rather than
setting them equal to $\pm 1$ by means of scaling). As said above, the
linearized solution (\ref{lin}) does not depend on the loss parameter $%
\alpha $ if it is small enough, $\alpha \ll \sigma \eta ^{2}$).

In the opposite case of $\sigma \eta ^{2}<0$ and small $\alpha \ll -\sigma
\eta ^{2}$, the linearized version of Eq. (\ref{LLE}) also supports a bound
state, which, however, is a loosely localized one, unlike the tightly bound
state (\ref{lin}):%
\begin{equation}
u_{\mathrm{lin}}^{\mathrm{(loose)}}\approx -\frac{\epsilon }{\sqrt{-2\sigma
\eta ^{2}}}\exp \left( \mp i\frac{\chi }{2}-\left( \frac{\alpha }{\sqrt{%
-2\sigma \eta ^{2}}}+i\sqrt{-2\sigma \eta ^{2}}\right) \left\vert x\pm \frac{%
l}{2}\right\vert \right) ,  \label{loose}
\end{equation}%
whose power is much larger than that given by Eq. (\ref{Plin}), as is
proportional to $\alpha ^{-1}$:%
\begin{equation}
P_{\mathrm{lin}}^{\mathrm{(loose)}}\approx \frac{\epsilon ^{2}}{\alpha \sqrt{%
-2\sigma \eta ^{2}}}.  \label{Ploose}
\end{equation}%
Finally, in the case of $\sigma =0$, when Eq. (\ref{LLE}) is linear by
definition, and the loss term is is the dominant one in it, the pinned
solution is%
\begin{equation}
u_{\mathrm{lin}}^{(\sigma =0)}=\frac{\epsilon \left( 1-i\right) }{2\sqrt{%
\alpha }}\exp \left( \mp i\frac{\chi }{2}-\sqrt{\alpha }\left( 1-i\right)
\left\vert x\pm \frac{l}{2}\right\vert \right) ,  \label{sigma=0}
\end{equation}%
the respective power being%
\begin{equation}
P_{\mathrm{lin}}^{(\sigma =0)}=\frac{\epsilon ^{2}}{2\alpha ^{3/2}}.
\label{Psigma=0}
\end{equation}

The value (\ref{Plin}) of the power pertains to the tightly bound linearized
solution (\ref{lin}) for the single delta-functional pump. In the case of
the pair of the pumps, defined as in Eq. (\ref{LLE}), the linearized
solution is a superposition of two fields (\ref{lin}) centered at $x=\pm l/2$%
. The corresponding expression for the total power is%
\begin{equation}
\left( P_{\mathrm{lin}}\right) _{\mathrm{total}}\approx \frac{\epsilon ^{2}}{%
\left( \sigma \eta ^{2}\right) ^{3/2}}\left[ \frac{1}{\sqrt{2}}+\left( \frac{%
1}{\sqrt{2}}+\sqrt{\sigma \eta ^{2}}l\right) e^{-\sqrt{2\sigma \eta ^{2}}%
l}\cos \chi \right] .  \label{Plin-tot}
\end{equation}%
In particular, in the limit of $l\rightarrow \infty $ Eq. (\ref{Plin-tot})
yields, naturally, the double value (\ref{Plin}) corresponding to the single
pump. In the opposite limit, $l=0$, Eq. (\ref{Plin-tot}) yields $\left( P_{%
\mathrm{lin}}\right) _{\mathrm{total}}\left( l=0\right) =\sqrt{2}\epsilon
^{2}\cos ^{2}\left( \chi /2\right) $, which corresponds to the double pump
in Eq. (\ref{LLE}).

In the limit of $\alpha \rightarrow 0$, which corresponds to Eq. (\ref%
{Plin-tot}), the analytical solution of the linearized LL equation (\ref{LLE}%
) does not produce the SB effect. Nevertheless, the solution of the
linearized equation exhibits SB if $\alpha $ is taken into regard. This is
easy to see in the limit of $\sigma =0$, when the field created by the
single HS is give by Eq. (\ref{sigma=0}). A straightforward calculation
produces the following expression for the local power profile created by the
two HSs with $\sigma =0$, separated by a finite distance $l$:%
\begin{gather}
\left\vert u_{\mathrm{lin}}^{(\sigma =0)}(x;l)\right\vert ^{2}=\frac{%
\epsilon ^{2}}{2\alpha }\left[ \sum_{+,-}\exp \left( -2\sqrt{\alpha }%
\left\vert x\pm \frac{l}{2}\right\vert \right) \right.   \notag \\
\left. +2\left\{
\begin{array}{c}
\exp \left( -2\sqrt{\alpha }x\right) \cos \left( l\sqrt{\alpha }-\chi
\right) ,~\mathrm{at~}x>+l/2, \\
\exp \left( -\sqrt{\alpha }l\right) \cos \left( 2\sqrt{\alpha }x-\chi
\right) ,~\mathrm{at~}|x|<l/2, \\
\exp \left( 2\sqrt{\alpha }x\right) \cos \left( l\sqrt{\alpha }+\chi \right)
,~\mathrm{at~}x<-l/2.%
\end{array}%
\right. \right]   \label{SBlinear}
\end{gather}%
It is obvious that the second term in Eq. (\ref{SBlinear}) indeed breaks the
symmetry between $x$ and $-x$ at $\chi \neq \pi n$ with integer $n$ (the
respective formal expression for SB (\ref{SSB}) is too cumbersome). This SB
mechanism is qualitatively similar to the one which explains SB in
collisions between solitons with a phase difference between them \cite{Khayk}. 

\begin{figure}[tb]
\centering \includegraphics[width=0.45\columnwidth]{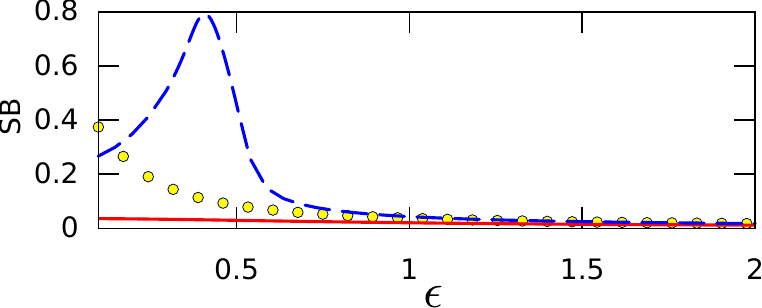}
\caption{The asymmetry ratio, SB, defined as per Eq. (\ref{SSB}),
versus $\epsilon $ for the self-repulsive system with $%
\sigma =-0.5$, $\chi =\pi /2$, $\alpha =0.1$, and $%
l=3$. The blue dashed line, yellow circles, and red solid line pertain to $%
\eta ^{2}=1$, $0$ and $-1$, respectively.}
\label{FIG_SSB_EP}
\end{figure}

\begin{figure}[tb]
\centering \includegraphics[width=0.9\textwidth]{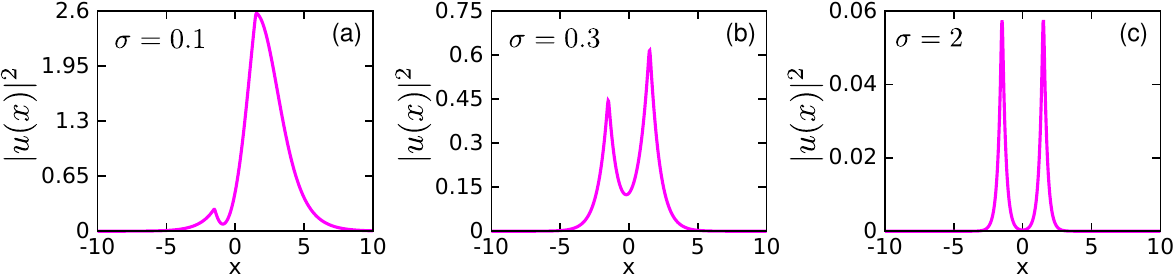}
\caption{Symmetric and asymmetric stationary profiles obtained numerically
from Eq. (\ref{LLE}) for different values of $\sigma>0$ with
$\chi=\pi/2$, $\epsilon=0.5$, $\eta^2=1$, $%
\alpha=0.1$, and $l=3$. }
\label{Fig_sig_pos}
\end{figure}

\begin{figure}[tb]
\centering \includegraphics[width=0.9\textwidth]{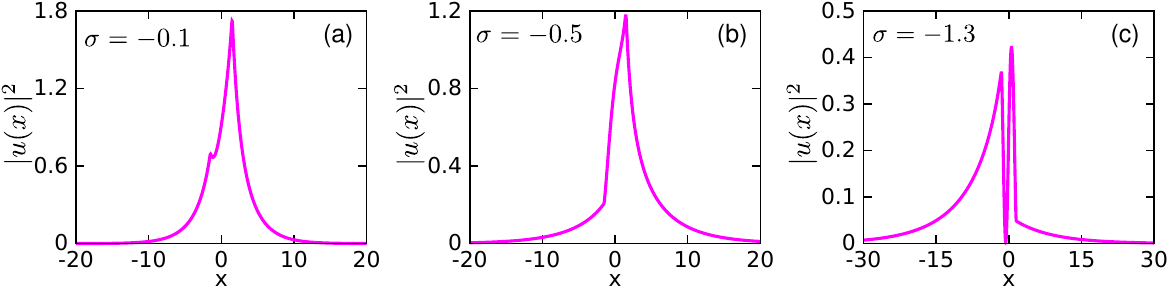}
\caption{The same as in Fig. \ref{Fig_sig_pos}, but for the system
with the self-repulsion ($\sigma<0$).}
\label{Fig_sig_neg}
\end{figure}

\begin{figure}[tb]
\centering \includegraphics[width=0.45\columnwidth]{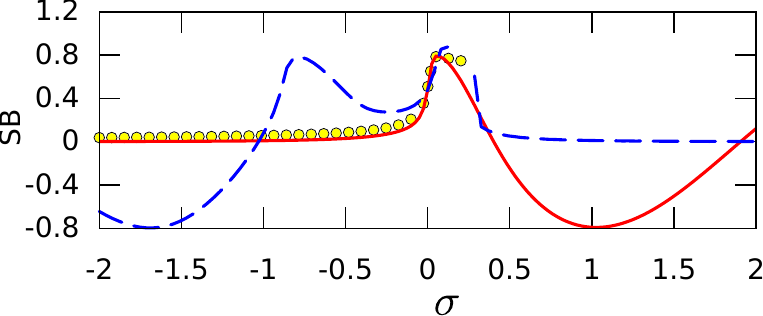}
\caption{The asymmetry ratio SB, defined as per Eq. (\ref{SSB}) vs. $%
\sigma$ for the system with self-repulsion and attraction with $%
\chi=\pi/2$, $\epsilon=0.5$, $\alpha=0.1$,
and $l=3$. The blue dashed line, yellow circles, and the red solid line
correspond to $\eta^2=1$, $0$ and $-1$, respectively. In the case of
the self-attraction, the system does not create stationary states in the
wide empty region.}
\label{FIG_SSB_SIG}
\end{figure}

\begin{figure}[tb]
\centering \includegraphics[width=0.9\textwidth]{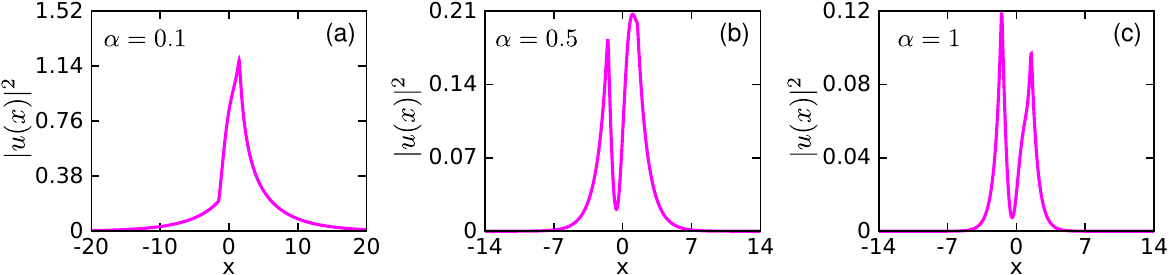}
\caption{Asymmetric stationary profiles obtained numerically from Eq. (%
\ref{LLE}) for different values of the loss coefficient $%
\alpha$ with $\chi=\pi/2$, $\epsilon=0.5$, $%
\eta^2=1$, $\sigma=-0.5$, and $l=3$.}
\label{Fig_alp}
\end{figure}

\begin{figure}[tb]
\centering \includegraphics[width=0.45\columnwidth]{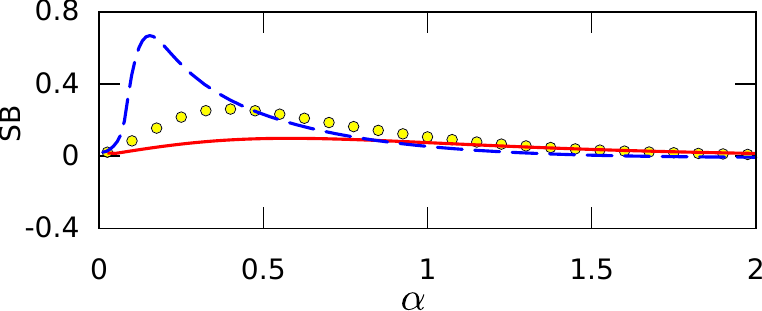}
\caption{The asymmetry ratio SB, defined as per Eq. (\ref{SSB}), vs.
$\alpha $ for the system with the self-repulsive nonlinearity, with $%
\sigma =-0.5$, $\chi =\pi /2$, $\epsilon %
=0.5 $, and $l=3$. The results in blue dashed line, yellow circles and red
solid line correspond to $\eta ^{2}=1$, $0$ and $-1$, respectively. }
\label{FIG_SSB_ALP}
\end{figure}

\section{Numerical results \label{sec:Numerical-results}}

\subsection{The shape and stability of the symmetric and asymmetric
stationary modes}

To produce stationary states admitted by Eq. (\ref{LLE}), we ran real-time
simulations of this equations, starting from the \textquotedblleft vacuum
state", $u(x,t=0)=0$, until the solution would converge to a stationary
state. Obviously, the states produced by this approach are stable.

We begin the analysis by studying the effect of $\eta ^{2}$ on the
stationary states. In Fig. \ref{Fig_et}, we present their profiles obtained
with $\alpha =0.1$, $\epsilon =0.5$, $\chi =\pi /2$, $l=3$ and $\sigma =-0.5$
(the self-defocusing sign of the nonlinearity), for $\eta ^{2}=+1$, $0$, and
$-1$. In this case, we observe that the asymmetry measure (\ref{SSB})
reduces with the decrease of $\eta ^{2}$. In particular, for $\eta ^{2}=+1$,
the resulting stationary state yields $\mathrm{SB}=0.46$, while for $\eta
^{2}=-1$, a small value $\mathrm{SB}=0.03$ is obtained. It is also observed
that the reduction of $\eta ^{2}$ leads to a decrease in the amplitude of
the stationary modes.

In Fig. \ref{Fig_chi} we address the dependence of the stationary states on
the phase difference $\chi $ between the pump beams. It is observed that, in
agreement with the above-mentioned argument, the states with $\chi =0$ (as
well as with $\chi =\pi n$, with integer $n$) keep the spatial symmetry. On
the other hand, the states obtained with the phase shift from interval $%
0<\chi <\pi $ exhibit asymmetry, the size of which depends on the other
parameters.

The dependence of the SB parameter on $\chi$ is presented in Fig. \ref%
{Fig_SSB_chi}, which exhibits the natural sinusoidal behavior with the
period of $2\pi $, which, naturally, includes positive and negative values.
The decrease in $\eta ^{2}$ leads to reduction of the amplitude of the
dependence, the minimum of which corresponds to $\eta ^{2}=-1$.

Figure \ref{Fig_ep} presents the comparison between power profiles, $%
\left\vert u(x)\right\vert ^{2}$, for the stationary states obtained with
different strengths $\epsilon $ of the pump beams. Naturally, the power
grows with the increase of the pump strength $\epsilon $. Furthermore, the
profiles drastically alter with the variation of $\epsilon $, directly
affecting the respective $\mathrm{SB}$ value. To complement this analysis,
in Fig. \ref{FIG_SSB_EP} we plot the relation between $\mathrm{SB}$ and $%
\epsilon $. It is observed that the increase of $\epsilon $ favors symmetric
profiles. This trend is explained by the above-mentioned fact that SB is
explained by the loss term $-\alpha u$ in Eq. (\ref{LLE}), while the
increase of $\epsilon $ makes the loss term relatively weaker. The same fact
explains the presence of nonvanishing SB at $\epsilon \rightarrow 0$, which
corresponds to the linear limit of Eq. (\ref{LLE}).

The increase of the self-interaction strength $\sigma $ promotes various
features of the profile of the stationary states. In general, the increase
of $\sigma >0$ (self-focusing) drives the profile narrowing. Other effects
are observed too, such as the reduction of $\mathrm{SB}$. Figure \ref%
{Fig_sig_pos} shows that the increase of $\sigma >0$ transforms the
asymmetric state into a narrow symmetric one with well-separated peaks,
which is explained by the same argument as presented above: the increase of $%
\sigma $ makes the asymmetry-inducing loss term relatively weaker.
Additionally, for certain values of $\epsilon $, which are detailed below,
stationary states are not supported in the in a self-focusing regime, while
this phenomenon does not occur in the self-defocusing case ($\sigma <0$).
Different stationary states obtained in this case are presented in Fig. \ref%
{Fig_sig_neg}. It is observed that the decrease of $\sigma $ (i.e., increase
of $|\sigma |$) promotes wider profiles, which affects the $\mathrm{SB}$.

In Fig. \ref{FIG_SSB_SIG} we present the effect of the self-interaction on $%
\mathrm{SB}$. For the profiles obtained with $\eta ^{2}=1$ and $-1$, the
behavior of the $\mathrm{SB}$ is quite complex, featuring both positive and
negative values of $\mathrm{SB}$. The dependences $\mathrm{SSB}(\sigma )$ is
opposite for $\eta ^{2}=1$ and $\eta ^{2}=-1$, as the symmetric states are
favored at $\sigma <0$ and $\sigma >0$ for $\eta ^{2}=-1$ and $1$,
respectively. In particular, the curve $\mathrm{SSB}(\sigma )$ for $\eta
^{2}=1$ features a discontinuity at $\sigma >0$ region, where stationary
profiles actually do not exist (were not found). For $\eta ^{2}=0$, the
value of $\mathrm{SSB}(\sigma )$ is very small at $\sigma <0$. In the
self-defocusing regime, the curve $SSB(\sigma )$ for $\eta ^{2}=0$ is
limited at $\sigma <0.3$, beyond which stationary profiled were not found.

The effect of loss parameter $\alpha $ on the shape of the stationary states
is presented in Fig. \ref{Fig_alp}. It is observed that the increase in $%
\alpha $ leads to a decrease of the total power $P$. In general, the
increase in $\alpha $ reduces $\mathrm{SB}$. To detail this effect, we
explored the change of SB with the variation of $\alpha $. In Fig. \ref%
{FIG_SSB_ALP}, it is observed that, in agreement with the above analysis of
the solutions of the linearized equation (\ref{LLE}), SB\ vanishes at $%
\alpha =0$. At large value of $\alpha $ the stationary states tend to
gradually restore their symmetry. Considering different values of $\eta ^{2}$%
, it is again observed that the lowest values of $\mathrm{SB}$ correspond to
$\eta ^{2}=-1$.

\begin{figure}[tb]
\centering \includegraphics[width=0.85\columnwidth]{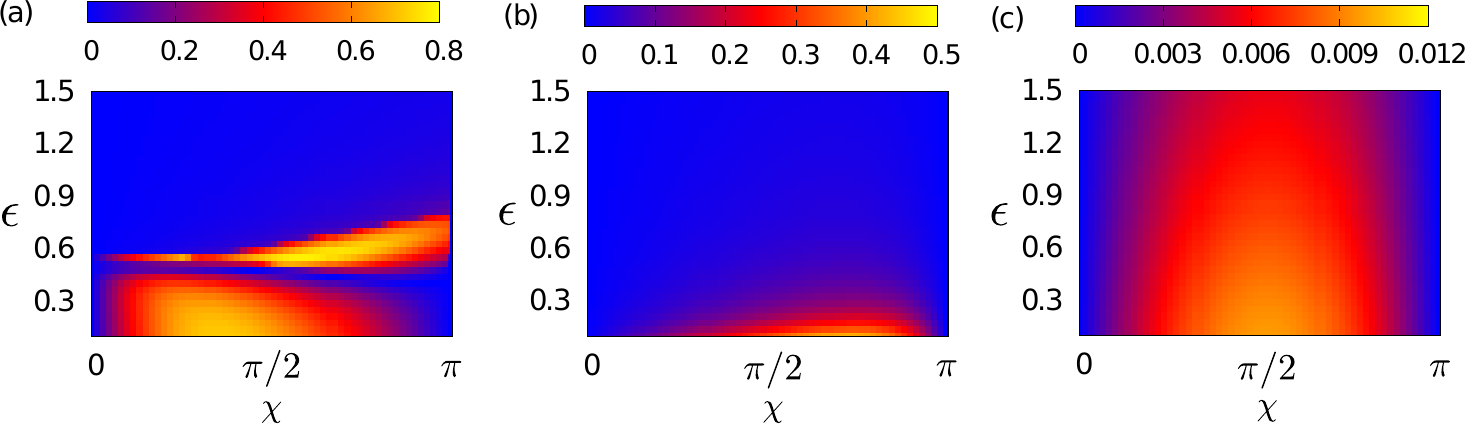}
\caption{The chart of the values of $|\mathrm{SB}|$ in the plane of the
phase shift $\chi$ and pump strength $\epsilon$, for (a) $%
\eta^2=1$, (b) $\eta^2=0$, and (c) $\eta^2=-1$. The
other parameters are $\alpha=0.1$, $\sigma=-1$, and $l=3$.}
\label{DIAG_XI_EP}
\end{figure}

\begin{figure}[tb]
\centering \includegraphics[width=0.85\textwidth]{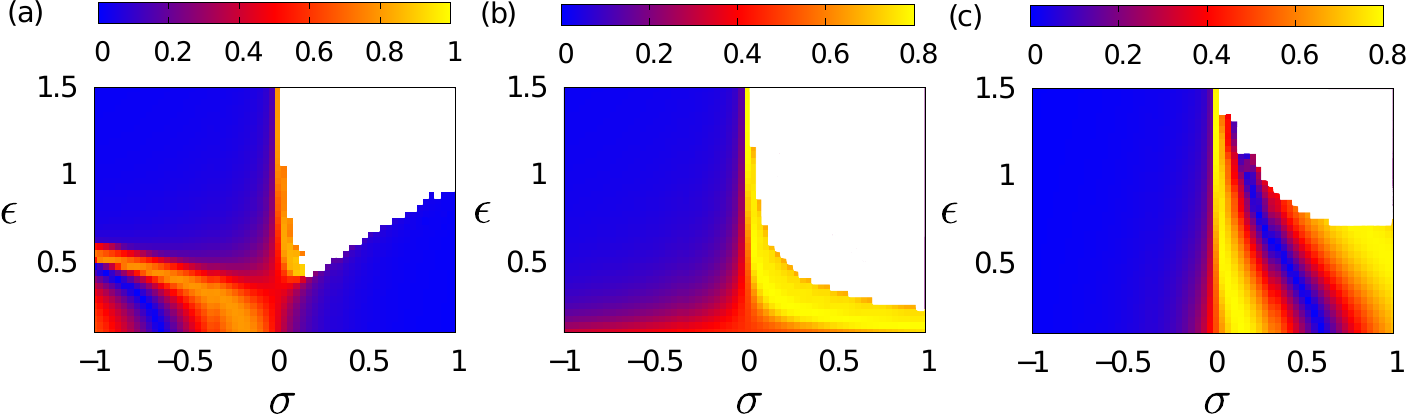}
\caption{The chart of the values of $|\mathrm{SB}|$ in the plane of the
self-interaction strength $\sigma $ and pump strength $%
\epsilon $, for (a) $\eta ^{2}=1$, (b) $\eta ^{2}=0$, and
(c) $\eta ^{2}=-1$. The other parameters are $\alpha =0.1$, $%
\chi =\pi /2$, and $l=3$. The stationary states were not
found in white areas.}
\label{DIAG_SIG_EP}
\end{figure}

\begin{figure}[tb]
\centering \includegraphics[width=0.85\columnwidth]{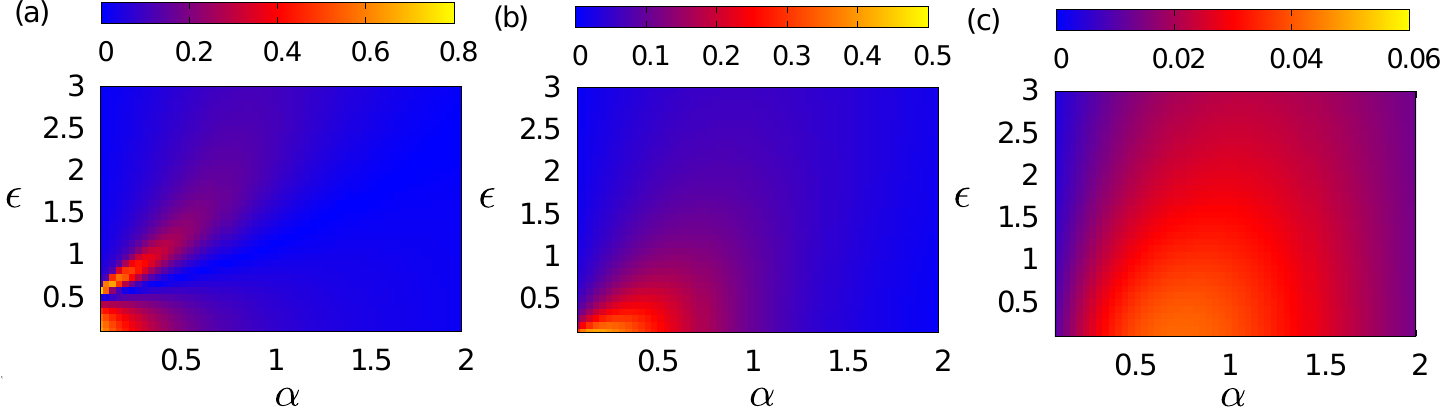}
\caption{The chart of the values of $|\mathrm{SB}|$ in the plane of the loss
coefficient $\alpha$ and pump strength $\epsilon$, for (a) $%
\eta^2=1$, (b) $\eta^2=0$, and (c) $\eta^2=-1$. The
other parameters are $\sigma=-1$, $\chi=\pi/2$, and $%
l=3$. }
\label{DIAG_ALP_EP}
\end{figure}

\begin{figure}[tb]
\centering \includegraphics[width=0.85\columnwidth]{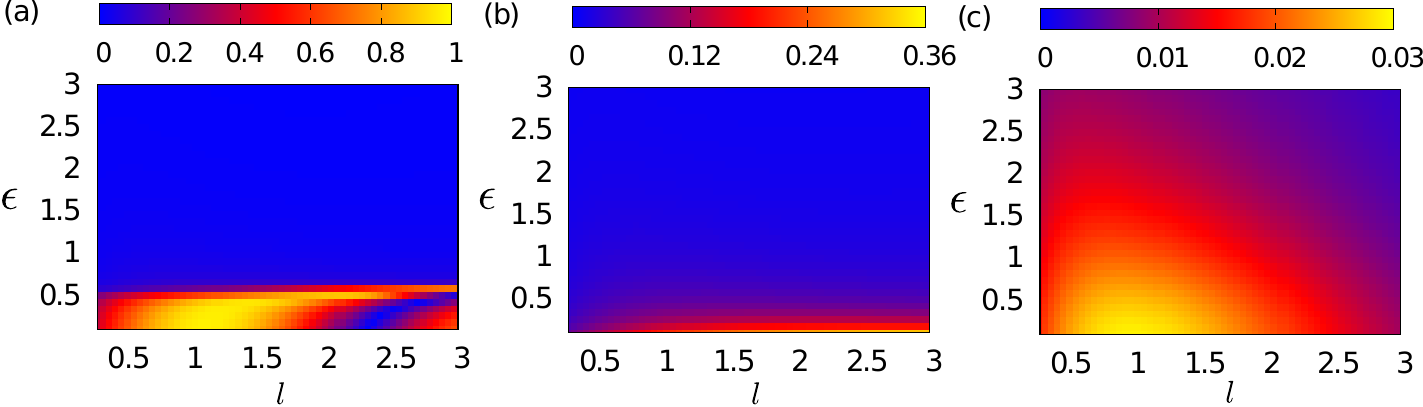}
\caption{The chart of the values of $|\mathrm{SB}|$ in the plane of the
separation between the delta-functional pumps $l$ and their strength $%
\epsilon $, for (a) $\eta ^{2}=1$, (b) $\eta ^{2}=0$%
, and (c) $\eta ^{2}=-1$. The other parameters are $\alpha %
=0.1$, $\chi =\pi /2$ and $\sigma =-1$.}
\label{DIAG_l_EP}
\end{figure}

\subsection{The summary of numerical findings}

To summarize the findings for SB, we produce color-coded charts for $\mathrm{%
|SB|}$ in parameter planes ($\chi $, $\epsilon $), ($\sigma $, $\epsilon $),
($\alpha $, $\epsilon $), and ($l$, $\epsilon $), plotted for $\eta ^{2}=1$,
$0$, and $-1$. In particular, the range of the phase-shift parameter $\chi $
is naturally confined to $0<\chi <\pi $.

First, in the chart plotted in the ($\chi ,\epsilon $) plane for $\eta ^{2}=1
$, it is observed that symmetric profiles are favored for intense pump
beams, i.e., for large values of $\epsilon $, as already mentioned above. As
shown in Fig. \ref{Fig_SSB_chi}, the largest values of $\mathrm{|SB|}$ are
found in a vicinity of $\chi =\pi /2$, in the weak-pump regime (small $%
\epsilon $). For $\eta ^{2}=0$, the chart demonstrates strong reduction of
the SSB area. In the chart for $\eta ^{2}=-1$, strong attenuation of SB is
observed in comparison to the cases of $\eta ^{2}=1$ and $0$. Note that, in
all the cases, only symmetric profiles are found for $\chi =0$ and $\pi $,
in agreement with the above analysis. These results are displayed in Fig. %
\ref{DIAG_XI_EP}.

In Fig. \ref{DIAG_SIG_EP}, we present the SB charts in the plane of ($\sigma
$, $\epsilon $). In all the three panels, there are white areas at $\sigma
>0 $ (self-focusing) in which stationary profiles were not found. Further,
it is observed that the asymmetry is favored for weak pump beams in the
self-defocusing system with $\eta ^{2}=1$. In the case of $\eta ^{2}=0$, the
SB again occurs mainly for weak beam pumps. On the contrary, for $\eta
^{2}=-1$ SB is favored in the self-focusing system. As stressed above, the
results demonstrate the persistence of SB in the linearized system, with $%
\sigma =0$.

The SB charts in the plane of ($\alpha $, $\epsilon $) are shown in Fig. \ref%
{DIAG_ALP_EP}. For $\eta ^{2}=1$, SB occurs solely at $\alpha <1$. In the
case of $\eta ^{2}=0$, SB occurs for small values of the loss coefficient $%
\alpha $ in system with weak pump beams. For $\eta ^{2}=-1$, the SB
parameter is very small when compared to the results for $\eta ^{2}=1$ and $%
0 $.

Next, in Fig. \ref{DIAG_l_EP} we address the dependence of SB on distance $l$
between the pump beams. In all cases, the increase of $l$ leads to
attenuation of \ SB and consequently the obtaining of symmetric profiles.
This trend is easily explained by the fact that the asymmetric part of the
local-power pattern obtained in Eq. (\ref{SBlinear}) decays exponentially
with the increase of $l$.

\begin{figure}[tb]
\centering \includegraphics[width=0.6\textwidth]{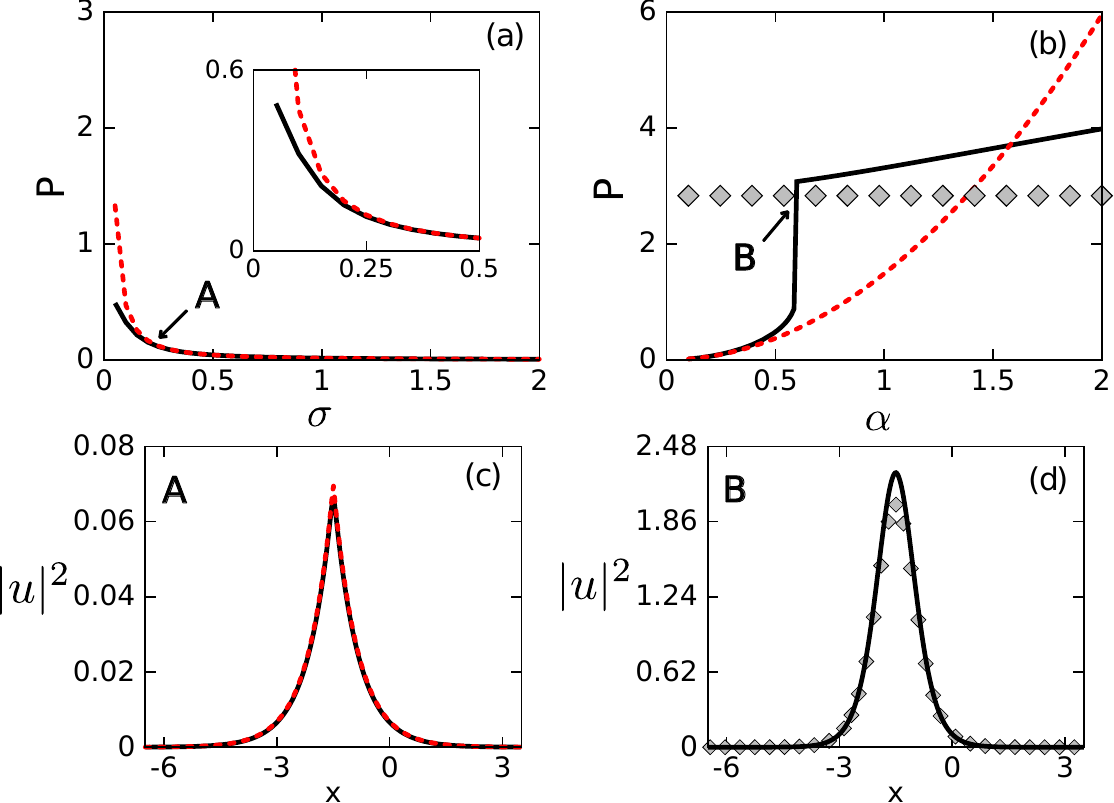}
\caption{The comparison between the approximate analytical solutions (%
\ref{lin}) and (\ref{sol}) and their numerical counterparts
found as stationary profiles of Eq. (\ref{LLE}) with the single
delta-function pump located at $x=-l/2\equiv -1.5$. Panels (a) and (b)
display the power $P$ of the stationary mode vs. $\sigma $ and $%
\alpha $, respectively. In (c) and (d), we present the comparison
between power profiles $|u(x)|^{2}$ of the numerically found solutions and
their approximate counterparts, corresponding to points \textbf{A}($%
\sigma =0.3$) and \textbf{B}($\alpha =0.61$), respectively. The
numerical solution is shown the black solid lines, while the approximations
produced by Eqs. (\ref{sol}) and (\ref{lin}) are depicted by
the chain of gray diamonds and the red dotted line, respectively. At point
\textbf{A}, solution (\ref{sol}) is irrelevant, therefore it is not
represented in panel (c). For the same reason, solution (\ref{lin})
is not displayed in (d). The results were obtained with $\epsilon %
=2.05\alpha $, $\eta ^{2}=1$ and $\alpha =0.1$ in
(a), and $\sigma =1$ in (b).}
\label{P_cub_P_lin}
\end{figure}

\begin{figure}[tb]
\label{P_cub_aprox}\centering \includegraphics[width=0.65%
\textwidth]{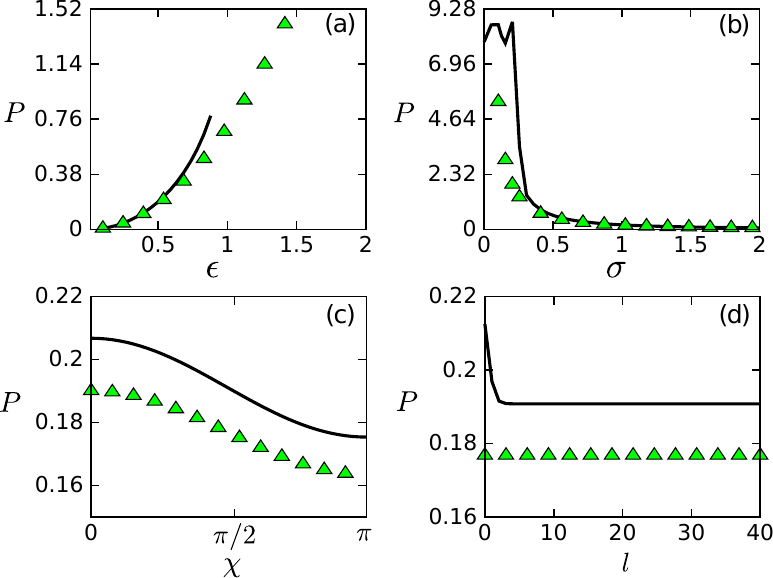}
\caption{Dependences of the power on parameters $\epsilon $, $%
\sigma $, $\chi $ and $l$. The power of the numerical
solutions of Eq. (\ref{LLE}) with $\eta ^{2}=1$ are
displayed by solid black lines, which are compared to the approximate
analytical expression $\left( P_{\mathrm{lin}}\right) _{\mathrm{total}}$, as
given by Eq. (\ref{Plin-tot}) and plotted by green triangles. The
other parameters are: (a) $\sigma =1$, $\alpha =0.1$, $%
\chi =\pi /2$ and $l=3$; (b) $\epsilon =0.5$, $%
\alpha =0.1$, $\chi =\pi /2$ and $l=3$; (c) $%
\sigma =1$, $\epsilon =0.5$, $\alpha =0.1$ and $l=3$; and
(d) $\sigma =1$, $\epsilon =0.5$, $\alpha =0.1$ and $%
\chi =\pi /2$.}
\end{figure}

\begin{figure}[tb]
\centering \includegraphics[width=0.9%
\textwidth]{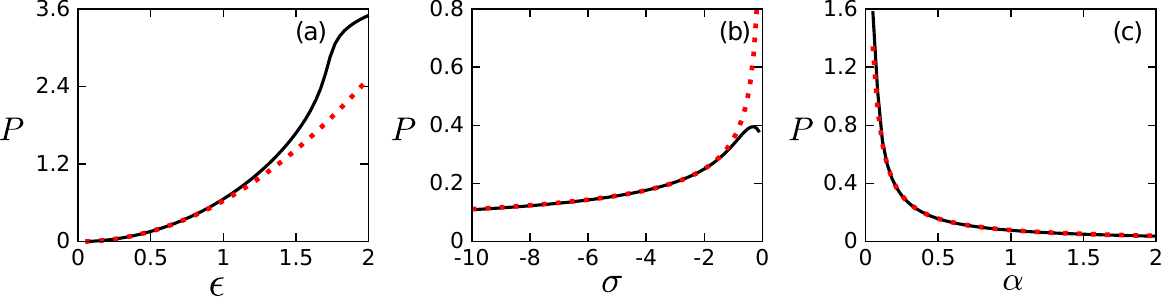}
\caption{Dependences of the power on parameters $\epsilon$, $%
\sigma $, and $\alpha$. The power of the numerical
solutions of Eq. (\ref{LLE}) with the single delta-function located at $l/2$ are
displayed by solid black lines, which are compared to the approximate
analytical expression $P_{\mathrm{lin}}^{\mathrm{(loose)}}$, as
given by Eq. (\ref{Ploose}) and plotted by red dotted lines. The
other parameters are: $\eta^2=1$ and $l=3$ (a) $\alpha =0.5$ and $\sigma =-5$; (b) $\epsilon =0.5$ and $\alpha =0.5$; (c) $\sigma =-5$ and $\epsilon =0.5$.}
\label{Fig_Ploose}
\end{figure}

\begin{figure}[tb]
\centering \includegraphics[width=0.6\textwidth]{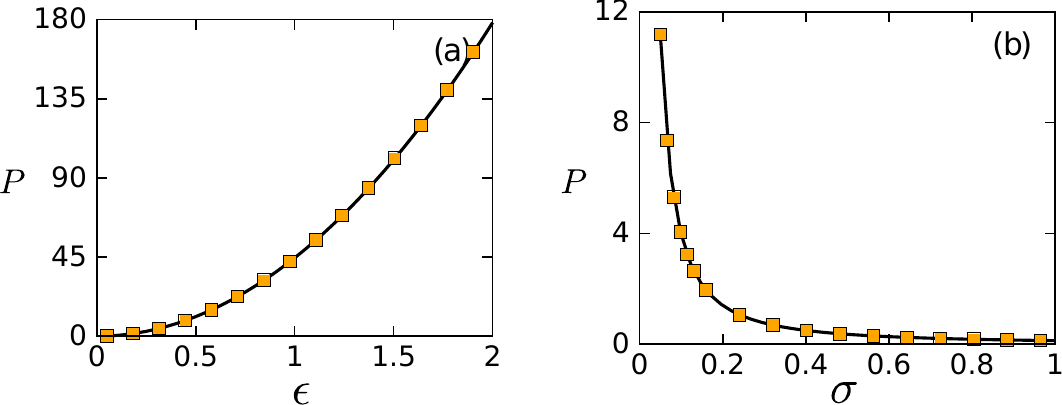}
\caption{Dependences of the power on parameters $\epsilon $ and $\alpha$. The power of the numerical
solutions of Eq. (\ref{LLE}) with the single delta-function are
displayed by solid black lines, which are compared to the approximate
analytical expression $P_{\mathrm{lin}}^{(\sigma =0)}$, as
given by Eq. (\ref{Psigma=0}) and plotted by orange squares. The
other parameters are: $\sigma = 0$, $\chi=\pi/2$ and $\l =3$ (a) $\alpha=0.05$; and (b) $\epsilon = 0.5$.}
\label{P_cub_P_lin_Eq13}
\end{figure}

\begin{figure}[tb]
\centering \includegraphics[width=0.6\textwidth]{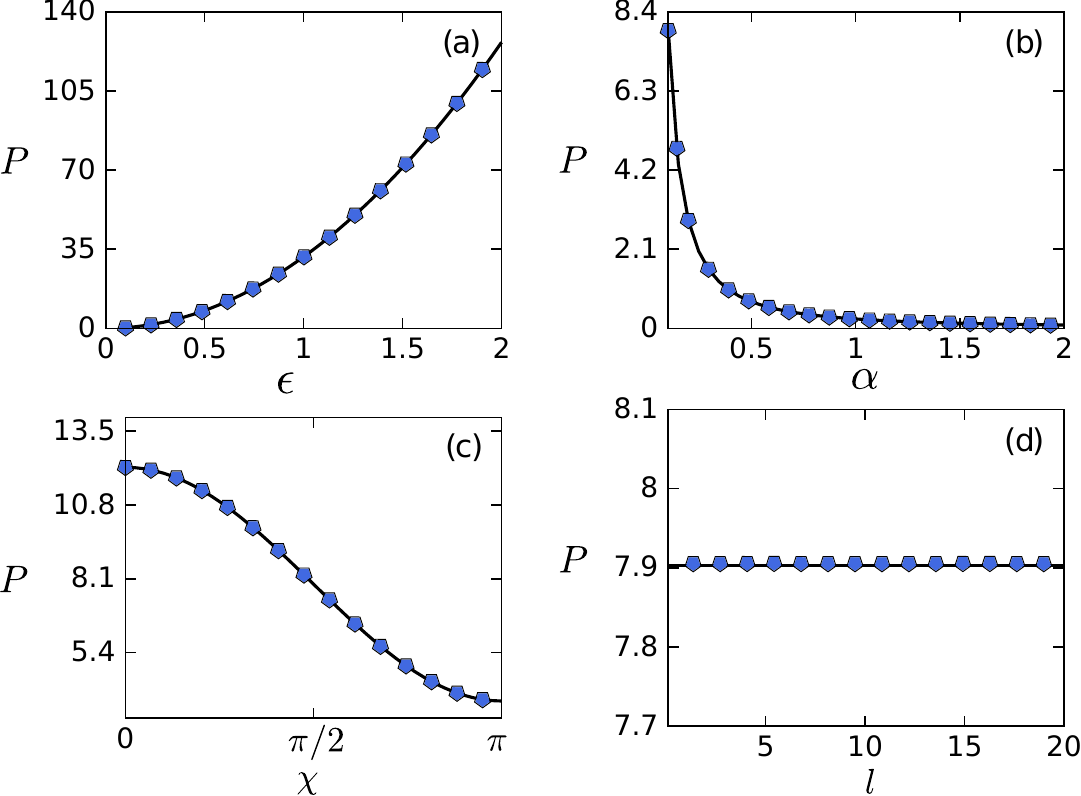}
\caption{Dependences of the power on parameters $\epsilon $, $%
\alpha $, $\chi $ and $l$. The power of the numerical
solutions of Eq. (\ref{LLE}) with $\eta ^{2}=1$ and $\sigma=0$ are
displayed by solid black lines, which are compared with the power of the approximate analytical solution $\left\vert u_{\mathrm{lin}}^{(\sigma =0)}(x;l)\right\vert ^{2}$, as
given by Eq. (\ref{SBlinear}) and plotted by blue pentagons. The
other parameters are: (a) $\sigma =1$, $\alpha =0.1$, $%
\chi =\pi /2$ and $l=3$; (b) $\epsilon =0.5$, $%
\alpha =0.1$, $\chi =\pi /2$ and $l=3$; (c) $%
\sigma =1$, $\epsilon =0.5$, $\alpha =0.1$ and $l=3$; and
(d) $\sigma =1$, $\epsilon =0.5$, $\alpha =0.1$ and $%
\chi =\pi /2$.}
\label{P_cub_P_lin_COMP}
\end{figure}

\begin{figure}[tb]
\centering \includegraphics[width=0.6\textwidth]{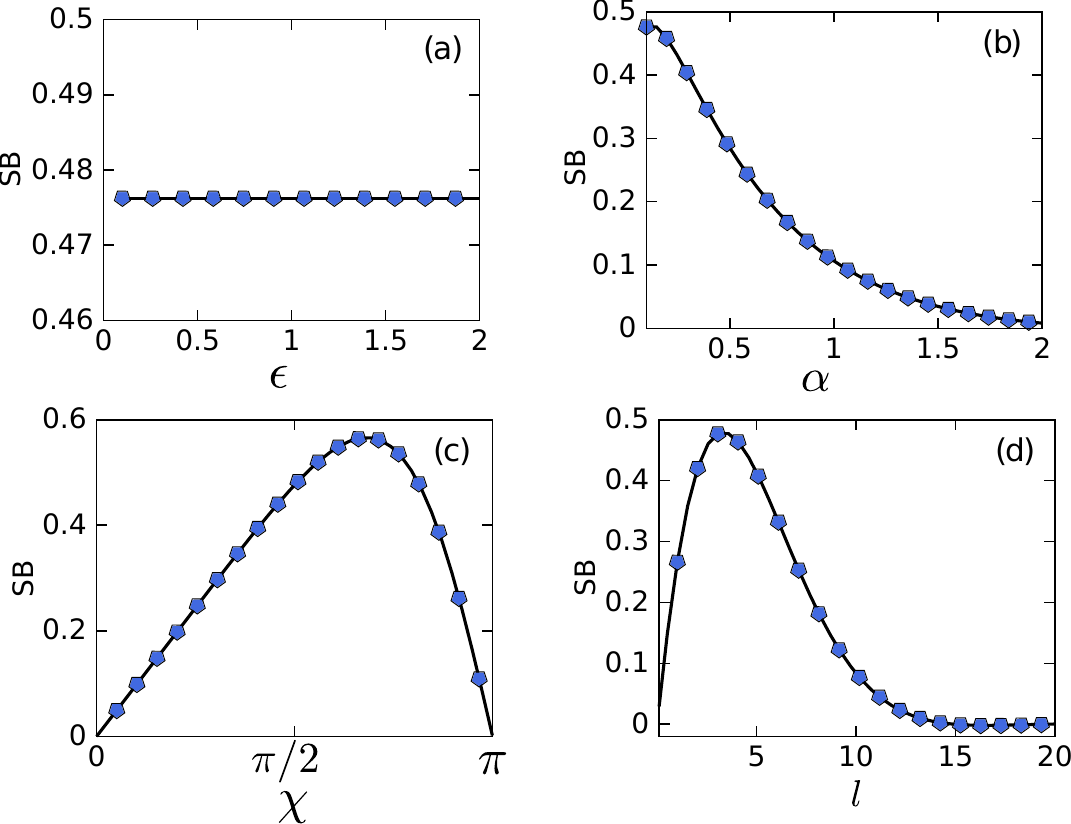}
\caption{The same as Fig. \ref{P_cub_P_lin_COMP}, but considering the symmetry breaking (SB), given by Eq. (\ref{SSB}).}
\label{P_cub_P_lin_COMP_SB}
\end{figure}

\subsection{Comparison of analytical and numerical results}

It is relevant to check the accuracy of the above-mentioned approximate
analytical solutions (\ref{sol}) and (\ref{lin}) . To this end, we use
stationary profiles produced by the numerical solution of Eq. (\ref{LLE}),
which includes the single delta-function pump. In Figs. \ref{P_cub_P_lin}(a)
and \ref{P_cub_P_lin}(b), we present the comparison of the analytically
predicted and numerically found dependences of power $P$ on nonlinearity
strength $\sigma $ and loss coefficient $\alpha $, respectively. In this
figure, the pump strength is taken as $\epsilon =2.05\alpha $, to consider
the situation close to the existence threshold (\ref{thr}) for the
approximate solution (\ref{sol}).

It is seen in Fig. \ref{P_cub_P_lin}(a) that the small-amplitude analytical
solution (\ref{lin}) produces the result which is close to its numerically
found counterpart, both being virtually identical at $\sigma >0.2$. Further,
Fig. \ref{P_cub_P_lin}(b) demonstrates that the same approximate solution (%
\ref{lin}) predicts the $P(\alpha )$ dependence which is close to its
numerical counterpart at $\alpha <0.5$, exhibiting strong discrepancy at
large values of $\alpha $.

The soliton-like approximate solution (\ref{sol}) has the constant ($\alpha $%
-independent) power, $P=2\sqrt{2}$, which is shown by the horizontal chain
of diamonds in Fig. \ref{P_cub_P_lin}(b). It is seen in this figure that the
soliton approximation is irrelevant for small values of $\alpha $, but at $%
\alpha =0.6$ the numerical solution suddenly makes a jump to a value close
to this constant, differing from it by $8.8\%$. At larger values of $\alpha $%
, the power of the numerical solution demonstrates a slow growth, so that
the analytical prediction, based on the soliton ansatz, remains a roughly
relevant one. The comparison between the profiles of the approximate
solutions, given by expressions (\ref{sol}) and (\ref{lin}), and their
numerically found counterparts at points \textbf{A} and \textbf{B} are shown
in Figs. \ref{P_cub_P_lin}(c) and \ref{P_cub_P_lin}(d), respectively.

The above comparison is performed for the LL equation with the single HS. In
the case of Eq. (\ref{LLE}) with the HS pair, the analytical prediction for
the power of small-amplitude (linear) stationary state was derived above in
the form of Eq. (\ref{Plin-tot}). It is compared to the power of the
numerically found symmetric solution in Fig. \ref{P_cub_aprox}. In Fig. \ref%
{P_cub_aprox}(a), we observe that the analytical prediction is close to the
numerical results at $\epsilon <0.88$, while at $\epsilon >0.88$ stationary
numerical solutions were not found. Other panels of Fig. \ref{P_cub_aprox}
demonstrate that the analytical approximation (\ref{Plin-tot}) provides a
relatively accurate prediction for the parameter dependences of the power.
In particular, the values of $P$ in the dependences $P(\chi )$ and $P(l)$,
displayed in panels (c) and (d), differ from their numerically produced
counterparts by $\approx 8\%$.

In Fig. \ref{Fig_Ploose}, we illustrate the dependencies of power on the parameters $\epsilon$, $\sigma$, and $\alpha$. This analysis considers the power of the numerical solutions derived from Eq. (\ref{LLE}), with a single delta function located at $l/2$, compared to the approximate analytical expression $P_{\mathrm{lin}}^{\mathrm{(loose)}}$, as defined by Eq. (\ref{Ploose}). Specifically, from Fig. \ref{Fig_Ploose}(a), it can be observed that the linear approximation holds valid for small values of $\epsilon$. Conversely, Fig. \ref{Fig_Ploose}(b) demonstrates that as the absolute value of negative parameter $\sigma$ increases, the fit of the linear approximation to the complete result improves. The final panel, Fig. \ref{Fig_Ploose}(c), illustrates that varying the damping parameter $\alpha$ - with $\sigma = -5$ and $\epsilon = 0.5$ held constant - consistently yields accurate results for the linear approximation.

Subsequently, we present  in Fig. \ref{P_cub_P_lin_Eq13} the behavior of power as a function of $\epsilon$ and $\alpha$ for the approximate analytical expression $P_{lin}^{(\sigma = 0)}$, given by Eq. (\ref{Psigma=0}), compared to the results obtained from the complete equation (Eq. (\ref{LLE})), with a single delta function located at $l/2$. Notably, for $\sigma = 0$, the approximate solution exhibits similar behavior to the results obtained numerically from Eq. (\ref{LLE}).

In the subsequent analyses, we investigate the approximate solutions for the case of a pair of pumps, as defined in Eq. (\ref{LLE}). Here, the linearized solution is a superposition of two fields centered at $x = \pm l/2$, as described by Eq. (\ref{SBlinear}). In this context, Fig. \ref{P_cub_P_lin_COMP} presents some examples of the power dependence on the system parameters when $\sigma = 0$. It can be observed once again that the power behavior of the approximate solution (Eq. (\ref{Plin-tot})) closely matches that produced by the numerical simulations of Eq. (\ref{LLE}), demonstrating the robustness of the solution.

The symmetry breaking, as characterized by the function given in Eq. (\ref{SSB}), was also analyzed for this approximate solution. In Fig. \ref{P_cub_P_lin_COMP_SB}, we display the behavior of the parameter related to the symmetry breaking of the approximate solution, given by Eq. (\ref{SBlinear}), alongside its numerical counterpart obtained from direct simulations of Eq. (\ref{LLE}). The behavior of both solutions is consistent, as evidenced by the panels in Fig. \ref{P_cub_P_lin_COMP_SB}, which highlight the behavior for various values of the system parameters.

\section{The double-HS system with the quintic nonlinearity}

The quintic self-focusing nonlinearity occurs in many optical materials \cite%
{Herve,Soumendu,Cid}, including those which may feature it in a nearly
\textquotedblleft pure" form, while the cubic term is negligible \cite{Cid2}%
. This fact suggests a possibility to introduce the LL equation with the
quintic nonlinear term \cite{Kozyreff,LL-quintic,Shatrughna}. In the present
context, we consider the quintic LL equation including the double HS,%
\begin{equation}
\frac{\partial u}{\partial t}=-\alpha u+\frac{i}{2}\frac{\partial ^{2}u}{%
\partial x^{2}}+i\sigma \left( \left\vert u\right\vert ^{4}-\eta ^{4}\right)
u+\epsilon \left[ e^{-i\chi /2}\delta \left( x+\frac{l}{2}\right) +e^{i\chi
/2}\delta \left( x-\frac{l}{2}\right) \right] ,  \label{LLEquint}
\end{equation}%
with $\sigma >0$ and $\eta ^{2}>0.$ cf. its cubic counterpart (\ref{LLE}).

On the other hand, a well-known fact is that the quintic self-focusing in 1D
waveguides gives rise to the \textit{critical collapse}; accordingly, the
family of the \textit{Townes solitons} (TSs) produced by the 1D\ NLS
equation, are completely unstable \cite{Salerno}. TSs are represented by the
following exact solution to Eq. (\ref{LLEquint}), with $\alpha =\epsilon =0$:
\begin{equation}
u_{\mathrm{TS}}(x)=\frac{3^{1/4}\eta }{\sqrt{\cosh \left( 2\sqrt{2\sigma }%
\eta ^{2}x\right) }}.  \label{Townes}
\end{equation}%
Its power is a constant which does not depend on real mismatch $\eta $,
\textit{viz}., $P_{\mathrm{TS}}=\left( \pi /2\right) \sqrt{3/\left( 2\sigma
\right) }$ (the degeneracy of the TS family, i.e., the independence of the
power on the phase mismatch, is a fundamental property of TSs \cite%
{Berge,Fibich}).

\begin{figure}[tb]
\centering \includegraphics[width=0.9\textwidth]{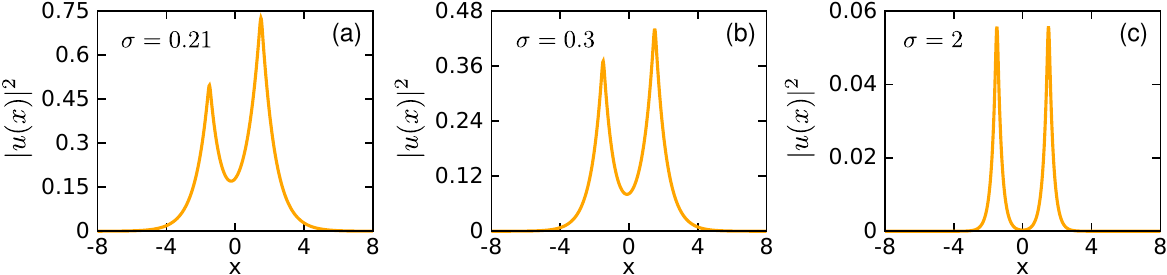}
\caption{Profiles of stable stationary states produced by the numerical
solution of the quintic LL model (\ref{LLEquint}) for different
values of the quintic coefficient $\sigma >0$. The other parameters
are $\alpha =0.1$, $\chi =\pi /2$, $\eta %
^{4}=1$, $\epsilon =0.5$ and $l=3$.}
\label{Fig_quint_perf}
\end{figure}

The instability of the TSs created by the quintic self-focusing suggests a
challenging possibility to check if the LL model with the quintic term can
support \emph{stable} localized modes pinned to an HS, or to the double HS,
as introduced in Eq. (\ref{LLEquint}).%
Following the method used above, we performed numerical solutions of Eq. (%
\ref{LLEquint}) with the zero initial conditions. As a result, families of
stable symmetric and asymmetric bound states, pinned to the double HS, have
been found. Typical examples of the symmetric and asymmetric stationary
profiles are displayed in Fig. \ref{Fig_quint_perf}. In this figure, values
of parameters are chosen to be similar to those used in Fig. \ref%
{Fig_sig_pos}, to highlight differences between the stationary states
produced by the cubic and quintic LL equations. In this configuration,
stationary profiles are found only for a sufficiently string quintic
nonlinearity, \textit{viz}., with $\sigma \geq 0.21$.

\begin{figure}[tb]
\centering \includegraphics[width=0.9\textwidth]{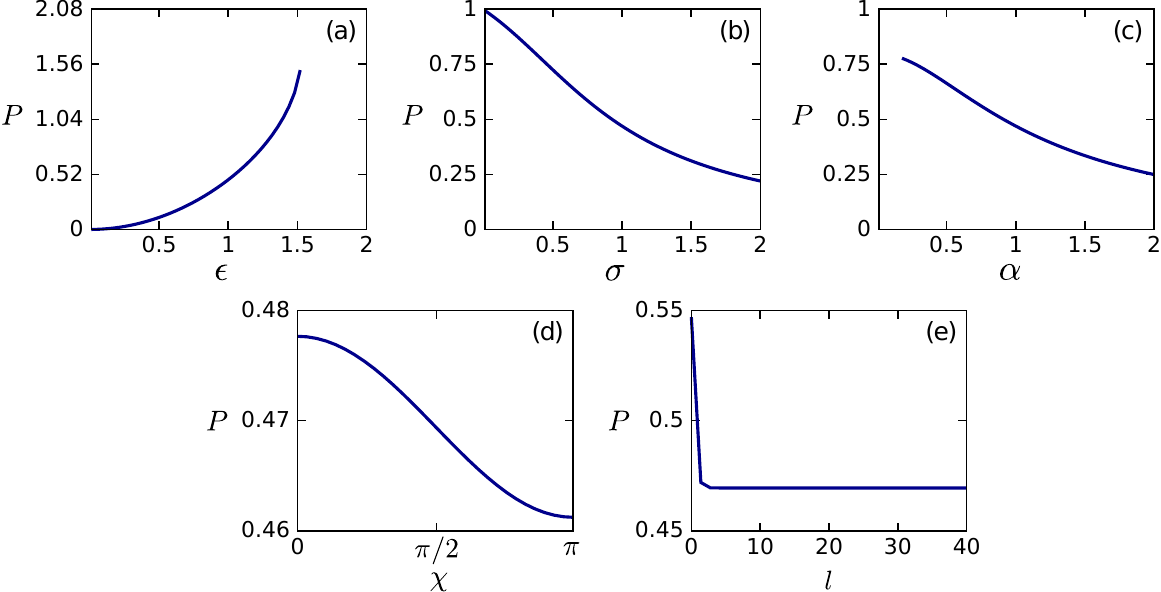}
\caption{Power of the stationary states produced by the quintic LL equation (%
\ref{LLEquint}) with $\eta ^{4}=1$ vs. parameters $%
\epsilon $, $\sigma $, $\alpha $, $\chi $ and $l$.
Other parameters are (a) $\alpha =1$, $\sigma =1$, $%
\chi =\pi /2$, and $l=3$; (b) $\epsilon =1$, $\alpha %
=1$, $\chi =\pi /2$, and $l=3$; (c) $\epsilon =1$, $%
\sigma =1$, $\chi =\pi /2$, and $l=3$; (d) $%
\epsilon =1$, $\sigma =1$, $\alpha =1$, and $l=3$; (e) $%
\epsilon =1$, $\sigma =1$, $\alpha =1$, and $%
\chi =\pi /2$. }
\label{Fig_quint_norm}
\end{figure}

\begin{figure}[tb]
\centering \includegraphics[width=0.85\columnwidth]{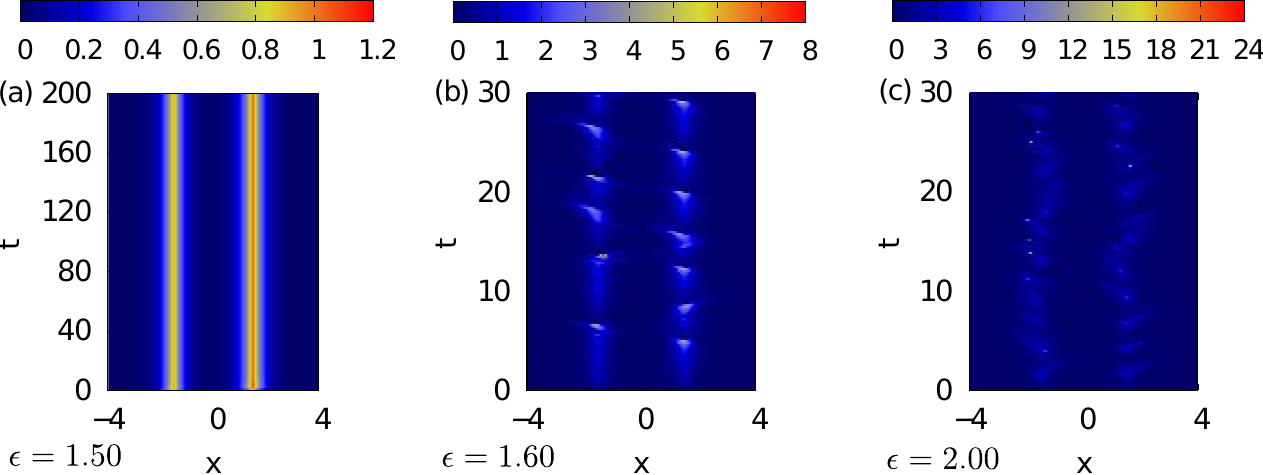}
\caption{(a) A stable bound state produced by the quintic LL equation (%
\ref{LLEquint}) with the pump strength $\epsilon =1.50.$
(b,c) The increase of $\epsilon $ to values $1.60$ and $2.00$ leads,
severally, to the onset of moderate and strong instability of the bound
state. Other parameters are $\eta ^{4}=1$, $\alpha =1$, $%
\sigma =1$, $\chi =\pi /2$, and $l=3$.}
\label{PERF_INST_EP}
\end{figure}

The families of stable stationary states produced by the quintic LL model (%
\ref{LLEquint}) are characterized, in Fig. \ref{Fig_quint_norm}, by
dependences of their power on parameters $\epsilon $, $\sigma $, $\alpha $, $%
\chi $, and $l$. Even for a very pump, with $\epsilon =0.01$, we find
stationary profiles with $P=4\times 10^{-5}$ in the case displayed in Fig. %
\ref{Fig_quint_norm}. In the same case, the increase of the pump strength
leads to the increase of power. The stationary stable state ceases to exist
at $\epsilon >1.52$. In latter case, the bound states develops oscillatory
instability, as shown in Fig. \ref{PERF_INST_EP}.

Figures \ref{Fig_quint_norm}(b) and (c) demonstrate that the power of the
stationary states decreases with the increase of the quintic and loss
coefficients, $\sigma $ and $\alpha $, respectively. In particular, the
stable stationary profiles are found, in Fig. \ref{Fig_quint_norm}(c), at $%
\alpha \geq 0.17$, while at $\alpha <0.17$ the bound states develop
oscillatory instability, similar to that displayed in Fig. \ref{PERF_INST_EP}%
.

\section{Conclusion \label{sec:Conclusion}}

We have introduced the one-dimensional LL (Lugiato-Lefever) equation
including the self-focusing or defocusing cubic nonlinearity and the
symmetric set of two tightly focused pumps (hot spots, HSs), approximated by
the delta-function, with the phase shift $\chi $ between them. Extensive
numerical simulations demonstrate the existence of the broad family of
stable bound states pinned to the double HS. The bound states are
characterized by their power and symmetry or asymmetry between the peaks
pinned to the individual HS. The boundary of the onset of SB (symmetry
breaking) is identified in the system's parameter space. The bound state
always remains symmetric for $\chi =0$ and $\pi $, as well as in the absence
of the loss term in the LL equation ($\alpha =0$). At $0<\chi <\pi $ and $%
\alpha >0$, SB takes place in the linearized and nonlinear (self-focusing or
defocusing) versions of the LL model alike, which is explained analytically.
Families of stable symmetric and asymmetric bound states pinned to the
double HS have been also found in the framework of the LL equation with the
quintic self-focusing, in spite of the full instability of the TSs (Townes
solitons) as solutions of the NLS equation with the same quintic
nonlinearity.

As extension of the work, Ref. \cite{Shatrughna} suggests a challenging
possibility to explore bound states, and the SB phenomenology in them, in
terms of the two-dimensional LL equation with local pumps (in particular,
vortical ones).

\section*{Acknowledgments}

W.B.C. acknowledges the financial support of the Brazilian agency CNPq
(grant \#306105/2022-5). This work was also performed as part of the
Brazilian National Institute of Science and Technology (INCT) for Quantum
Information (\#465469/2014-0). The work of S.K. and B.A.M. is supported, in
part, by the Israel Science Foundation through grant No. 1695/22.

\end{document}